\documentclass[12pt]{article}
\usepackage{epsfig}
\newcommand{\f}{\frac}
 
\def\be{\begin{equation}}
\def\ee{\end{equation}}
\def\beqna{\begin{eqnarray}}
\def\eeqna{\end{eqnarray}}
\def\bear{\begin{eqnarray}}
\def\eear{\end{eqnarray}}
\def\bearst{\begin{eqnarray*}}
\def\eearst{\end{eqnarray*}}
\begin{document}

\begin{center}
{\large\bf Shortcuts in a Dynamical Universe and the Horizon Problem}
\end{center}
\vspace{1ex}
\centerline{\large Elcio Abdalla and Adenauer Girardi Casali}
\begin{center}
{Instituto de Fisica, Universidade de S\~ao Paulo\\
C.P.66.318, CEP 05315-970, S\~ao Paulo, Brazil.}
\end{center}
\vspace{6ex}

\begin{abstract}

We consider the dynamics of a FRW brane in a purely AdS background,
from the point of view of the bulk, and explicitly construct the
geodesical behaviour of gravitational signs leaving and subsequently
returning to the brane. In comparison with photons following a geodesic 
inside the brane, we verify that shortcuts exist, though they are 
extremely small for today's Universe. However, we show that at 
times just before nucleosynthesis, if high redshifts were available,
the effect could be sufficiently large to  solve the horizon problem.
Assuming  an inflationary epoch in the brane evolution, we argue that
the influence of those signs trought the extra dimension in the causal
structure cannot be neglected. This effect could be relevant for probing the
extra dimension in inflationary scenarios.

\end{abstract}

\section{Introduction}
As it has been recently argued in \cite{veneziano}, although the standard 
model of particle physics is the uncontested theory of all
interactions down to distances of $10^{-17}$m, there are good reasons to
believe that new physics could be soon arising at the experimental
level. On the other hand, on the purely theoretical side, string theory
and further developments connected to it provide a background with immense
appeal in order to solve long-standing problems of theoretical high energy
physics. It is by now a widespread idea, from a general theoretical setup,
that the so-called M-theory \cite{polchinski} is a reasonable description
of our Universe: in the field theory limit, it is described by a solution 
of the (eventually 11-dimensional) Einstein equations with a cosmological 
constant, by means of a four dimensional membrane. In this picture only 
gravity survives in the higher dimensions, while the remaining matter and 
gauge interactions are typically four dimensional.

If the picture above is the one physically realized in nature, a very
large amount of new physics emerges. In particular, when membranes are
solutions of Einstein's equations and matter fields reside inside the brane,
the gravitational fields have to obey the Israel conditions
\cite{israel} at the sides of the brane. Thus, there is a
possibility that  gravitational fields propagating out of the brane
speed up, reaching farther distances as compared to
light propagating inside the brane, a scenario that for a resident of the
brane (such as ourselves) implies shortcuts \cite{Abdalla2} - \cite{Chung}.

 It is our aim in this work to further develop these ideas in
the case of a FRW brane Universe \cite{radion}. The subject was
developed until now from the point of view of the brane \cite{Abdalla}, where all
time dependence is embedded  in the bulk metric written in gaussian
coordinates. The price payed for the knownledge of the position of the
brane in the bulk is the complicated form of the bulk metric and,
consequently, the complicated behaviour of geodesics in the
bulk. However, if we treat the problem from the  point of view of the bulk, where the brane evolves in a non-trivial way in a static AdS
background, we can construct explicitly the causal structure of null
geodesics leaving and subsequently returning to the brane.  As it turns out,
shortcuts are common, although harmless at the present days (the delay is
vanishingly small), but could be large in the era before
nucleosynthesis if high redshits were available.

Moreover, one of the main goals of string theory nowadays is to 
prove itself able of coping with experimental evidences.  Branes have
been shown to be useful tools to understand the physics of strings and
M-theory \cite{Generalbranes}. As it has 
also recently been pointed out \cite{transdimen}, brane Universes such as 
the one described above could imply the existence of relics of the
extra dimensions in the cosmic microwave background. Unfortunately, recent
developments with  inflation guided by a scalar field in the
brane indicate that the consistency equation is
preserved \cite{inflation}.  In this
work however, we show that if inflation took part on the brane, the
causal structure is definitely changed by those gravitational
shortcuts, possibly leading to a non-usual period of causal evolution of
scales. This could be responsible for distinct  predictions in the
cosmic microwave background structure for inflationary models.

The paper is organized as follows. Chapter two provides a short
revision of the general setup adopted in this work: the view of the
bulk. In chapter three we discuss how the brane evolves in the background
and show the effects of the propagation of gravity trought the extra
dimension. Finally, in chapter 4 we explicitly construct the
geodesical behaviour of the shortcuts found in the earlier fase of the
evolution of the universe. A brief conclusion and discussion is also presented.

\section{The General Setup}

We consider a scenario where the bulk is a purely Anti-de-Sitter
space-time of the form \cite{ida},
\[
ds^{2} = h(a) dt^2 - \f{da^{2}}{h(a)} - a^2d\Sigma^2,
\]
with $ h(a) = k + \f{a^{2}}{l^2}$, $l\sim 0.1mm$ is the Randall-Sundrum
lenght scale \cite{RS} and $d\Sigma^2$ represents the metric of the three
dimensional spatial sections with  $k= 0\; , \;\pm 1$,
\[
d\Sigma^2 = \f{dr^{2}}{1-kr^2} + r^{2}[d\theta^{2} + \sin^{2}(\theta)d\phi^{2}].
\]

The brane is localized at $a_{b}(\tau)$, where $\tau$ is the proper time 
on the brane.
The unit vector normal to the brane is defined as (overdot and prime superscript denote differentiation
with respect to  $\tau$ and  $a$ respectively)
\be
n = \dot{a}_{b}(\tau)dt - \dot{t}(\tau)da\quad . \label{normal}
\ee
The normalization of $n$ implies the relation between the bulk time
$t$ and the brane time $\tau$,
\be
h(a_b)\dot{t}^{2} - \dot{a_b}^{2}h^{-1}(a_b) = 1
\label{nn}
\ee
and also  the usual FRW  expression for the distance
in the brane ,
\[
ds^2
= d\tau^{2} - a_b^{2}d\Sigma^{2}\quad .
\]

Aiming at the Israel conditions in the brane \cite{israel} we compute
the second fundamental form
\[
K_{i j} = e^{\alpha}_{i}e^{\nu}_{j}\nabla_{\alpha}n_{\nu}\quad ,
\]
\bear
K_{rr} &=& - \f{a_bh}{(1-kr^{2})}\dot{t} \quad , \label{krclasseII}\\
K_{\theta\theta} &=& - a_br^{2}h\dot{t}\quad
,\label{kthetaclasseII} \\
K_{\phi\phi} &=& - a_br^{2}\sin^{2}(\theta)h\dot{t}\quad
\label{kphiclasseII} \\
\eear
and 
\bear
K_{\tau\tau} &=&   \f{1}{h\dot{t}}\Bigl(\ddot{a}_b + \f{h'}{2} \Bigr)\quad .
\label{ktauclasseII}
\eear

With these results, using  the Israel conditions
for a $Z_{2}$ symmetric configuration \cite{israel}, 
\be
K_{ij} = \f{1}{2}\kappa_{(5)}^{2}\Bigl(S_{ij} - \f{1}{3}h_{ij}S\Bigr)\quad ,
\ee
we relate the
discontinuity in the second fundamental form through the brane and the
energy momentum tensor in the brane, $S_{ij}$. 
For an isotropic distribution of matter, as given by
\be
S_{ij} = \epsilon_T u_{i}u_{j} - p_T(h_{ij} - u_{i}u_{j})\quad ,
\ee
the following relations hold \cite{Kraus},\cite{BCG}, \cite{binetruy0}
\bearst
\f{d \epsilon_T}{d \tau} &=& - 3\f{\dot{a}_b}{a_b}(\epsilon_T + p_T)\quad 
\label{energy}
\eearst
and
\be
\f{\dot{a}_{b}^2}{a_{b}^2} + \f{h}{a_b^{2}} =
\f{\kappa_{(5)}^{4}\epsilon_T^{2}}{36}\quad .
\ee

Following \cite{Csaki2} and \cite{Cline}, we introduce an intrinsic
 non-dynamical energy density $\epsilon_0$ defined by means of $ \epsilon_T = \epsilon_0
 + \epsilon $, $p_{T} = -\epsilon_{0} + p$ where $\epsilon$ and $p$
 corresponds to the brane matter.  Thus the
 junction equations imply the usual energy conservation in the brane,
\bear
\f{d \epsilon}{d \tau} &=& - 3\f{\dot{a}_b}{a_b}(\epsilon + p)\quad 
\label{energy2}
\eear
and the  modified Friedmann equation \cite{binetruy},
\be
H^{2} = \Bigl(\f{\dot{a}_{b}}{a_{b}}\Bigr)^2 =
 \f{\Lambda_4}{3} + \f{1}{M_{Pl}^2}\Bigl(\f{\epsilon}{3} +
 \f{\epsilon^2}{6\epsilon_{0}}\Bigr)  -
 \f{k}{a_{b}^{2}}\quad , \label{Friedmann}
\ee
with the hierarchy
\be
M_{PL}^{-2} = \f{\kappa_{5}^{4}\epsilon_{0}}{6}\quad  \label{hierarquia}
\ee
and the cosmological constant in the brane
\be
\f{\Lambda_{4}}{3} = \Bigl(\f{\kappa_{5}^{4}\epsilon_{0}^{2}}{36} - \f{1}{l^{2}}\Bigr).
\ee

The present density of the Universe is  
\[
\epsilon(0) = \Omega_0 \epsilon_{c} = \Omega_{0} 3M_{PL}^{2}H_{0}^{2}\quad,
\]
where $\Omega_0$ is the ratio of the density and the critical density of
the Universe. Aiming at the energy conservation, the Friedmann equation can be
written as
\be
H^{2} = \f{\Lambda_4}{3} +\Omega_{0}
H_0^2\f{a_{b0}^{q}}{a_{b}^{q}}\Bigl(1  + \f{\Omega_{0}}{4(1+\Lambda_4l^2/3)}\f{L_{c}^{q}}{a_{b}^{q}}\Bigr) - \f{k}{a_b^{2}}\quad,
\label{Friedmann2}
\ee
where $ L_{c}^q = a_{b0}^{q}l^2H_0^{2}$.

We thus verify that there exist three phases in the evolution of the Universe.
When $a_{b}>>L_{c}$ the linear term in the energy density prevails in the
Friedmann equation, leading to the standard cosmology. Because
$H_0l\sim 10^{-29}$ this happens in a redshift of $a_{b0}/a_{b} \sim
10^{15}$,  much earlier than the nucleosynthesis.  For  $a_{b}<<L_{c}$
the Universe expands at a slower pace as compared to the standard model,
$a_b \propto \tau^{1/q}$, and the quadratic term is the prevailing one.
In an intermediate era where $a_{b} \sim L_{c}$, both phases coexist.

We see that all the cosmological information is already known:  the position of the brane in the extra dimension,
$a_{b}(\tau)$, is just the scale factor of the FRW metric and  the
junction conditions imply that the
cosmological evolution of the brane is obtained by the usual energy
conservation (\ref{energy2}) and the modified Friedmann equations
(\ref{Friedmann}).  However, in this work,  we are particularly interested in the evolution of the brane
with respect to the bulk. This relation can be obtained  using
(\ref{nn}) and transforming
from the time of the brane to the time of the bulk. Thus the position of the brane can also be treated as a
function of the bulk proper time satisfying the equation
\be
\f{d a_{b}}{dt} = \f{d a_{b}}{d\tau}\f{d \tau}{dt} = \dot{a}_{b}(\tau)
\f{h(a_{b})}{\sqrt{h(a_{b}) + \dot{a}_{b}(\tau)^2}}\quad . \label{bulk}
\label{timetime}
\ee

In the next section we solve the Friedmann equation (\ref{Friedmann2}) in order to
obtain the evolution of the brane inside the bulk, namely
(\ref{bulk}).

\section{The Brane evolving in the Bulk}

The main reason to study the evolution of the brane from the
point of view of the bulk is to simplify the analysis of gravitational
signs leaving and subsequently returning to the brane. In fact in the
static AdS backgound, the equation for a null geodesic in the bulk,
$a=a(t)$, is particularly simplified  \cite{Abdalla2},
\be
\f{\ddot{a}(t)}{a} + \f{\dot{a}^{2}(t)}{a^{2}}\Bigl(1 -
\f{3h'(a)a}{2h(a)}\Bigr) + \f{h(a)}{a}\Bigl(\f{h'(a)}{2} - \f{h}{a}\Bigr)=0\quad.
\label{nullgeodesic}
\ee

On the other hand, the evolution of the brane in the bulk,
$a=a_{b}(t)$,  at early times, $a_{b}<<L_{c}$, is dictated by
\[
 \f{d a_{b}}{d t} =  \f{h(a_{b})}{\sqrt{1 +
 \f{h(a_{b})}{\dot{a}_{b}(\tau)^2}}} = h(a_{b})\Bigl( 1 + \f{k +
 \f{a_{b}^{2}}{l^{2}}}{\f{\Omega_{0}^2}{4(1+l^2\Lambda_4)}
\f{L_c^{2q}}{l^2a_{b}^{2q-2}}}\Bigr)^{-1/2}\quad ,
\]
where we used the fact that the quadratic term in the energy prevails. Thus,
\[
 \f{d a_{b}}{d t} = h(a_{b})\Bigl( 1 +
 \f{4(1+l^2\Lambda_4)}{\Omega_{0}^2}\f{a_{b}^{2q}}{L_{c}^{2q}}(1 +
 kl^2/(a_b^2))\Bigr)^{-1/2}\quad ,
\]
and, for $a_{b}<<L_{c}$, since the observed cosmological constant is
at most $\Lambda_4 \sim H_0^{2}$, $\Lambda_4l^2<<1$ we have
\be
 \f{d a_{b}}{d t} \approx h(a_{b})\Bigl( 1 -
 \f{2(1+l^2\Lambda_4)}{\Omega_{0}^2}\f{a_{b}^{2q}}{L_{c}^{2q}}(1 + 
kl^2/(a_b^2))\Bigr)\quad . \label{primitivo}
\ee

Substituting the result for the evolution
of the brane, $\dot{a}(t) = h(a)$, in the geodesic equation (\ref{nullgeodesic}),
we verify that it is satisfied. Therefore,  the trajectory of the brane
differs from the trajectory of the null geodesic by a term of the order
$\Bigl(\f{a_{b}}{L_{c}}\Bigr)^{2q}$. 

This means that for $a_{b}<<L_{c}$, the trajectory of the brane in the
bulk is governed by
\be
\f{da_{b}(t)}{dt} = k + \f{a_{b}^{2}}{l^{2}}\quad . \label{branageo}
\ee

Thus, if $k=0$,
\be
a_b(t) = \f{a_b(0)l^{2}}{l^{2} - a_b(0)t}
\label{abtran}
\ee
and, if $k=-1$,
\be
a_b(t) = \f{2l(l+a_b(0))}{e^{2t/l}(l-a_b(0)) + l + a_b(0)} - l\quad.
\ee

In this last situation, if the Universe begins under the
Randall-Sundrum scale, $a(0)<l$, it will recolapse to the singularity in a
finite time,
\[
t = \f{l}{2}\ln \Bigl(\f{l+a_b(0)}{l-a_b(0)}\Bigr)\quad.
\] 
Indeed, there is an event horizon when $a=l$ if $k=-1$.

In the case of an eliptic Universe,
\be
a_{b}(t) = l\tan\Bigl(\f{t}{l} + \tan^{-1}\Bigl(\f{a_b(0)}{l}\Bigr)\Bigr)\quad.
\ee
Starting at the singularity
$t_{0}=\tau_{0}=a_{b}(t_0)=0$,  we have
\be
a_{b}(t) = l\tan\Bigl(\f{t}{l}\Bigr)\quad .
\ee 
Note that the evolution of the brane in the bulk is linear near the
initial singularity ($a(t)\sim t$ for $t<<l$), diverging at the critical
time $t_{c} = \f{\pi}{2}l$. In fact, the behaviour of all solutions is
similar near the critical time
\bear
t_{c} &=& \f{l^2}{a_b(0)}\quad;\nonumber\\
t_{c} &=& \f{l}{2}\ln
\Bigl(\f{l+a_b(0)}{a_b(0)-l}\Bigr)\quad;\nonumber \\
t_{c} &=& \f{\pi}{2}l - l\tan^{-1}\Bigl(\f{a_b(0)}{l}\Bigr)
\eear
for $k=0,\: -1,\: +1$ respectively.

\begin{figure*}[htb!]
\vspace{-2cm}
\begin{center}
\leavevmode
\begin{eqnarray}
\epsfxsize= 9.0truecm \rotatebox{-90}
{\epsfbox{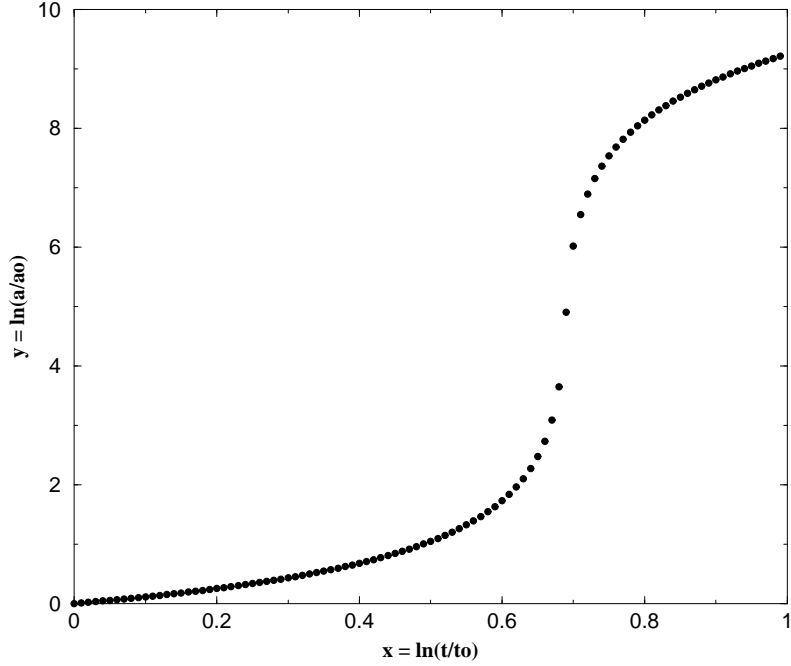}}\nonumber
\end{eqnarray}
\caption{Evolution of the brane in the bulk at the intermediate epoch
$a_b\sim 0.01L_{c}$ until $a_b \sim 200 L_{c}$, radiation dominated era 
(RDE), $\Omega_0=2$.}
\label{horizonte2}
\end{center}
\end{figure*}

As we approach the critical time
$a_{b}(t)$ increases quickly. When $a_{b}(t)\sim L_{c}$ equation
(\ref{branageo}) is no longer valid, and the trajectory of the brane is no
longer a geodesic. Thus, for a very short period, from the point of view
of the bulk the brane undergoes a phase transition. In figure
\ref{horizonte2} this behaviour is shown in terms of the numerical
solution of eq. (\ref{bulk}) for a radiation dominated eliptic Universe.

\begin{figure*}[htb!]
\vspace{-2cm}
\begin{center}
\leavevmode
\begin{eqnarray}
\epsfxsize= 9.0truecm \rotatebox{-90}
{\epsfbox{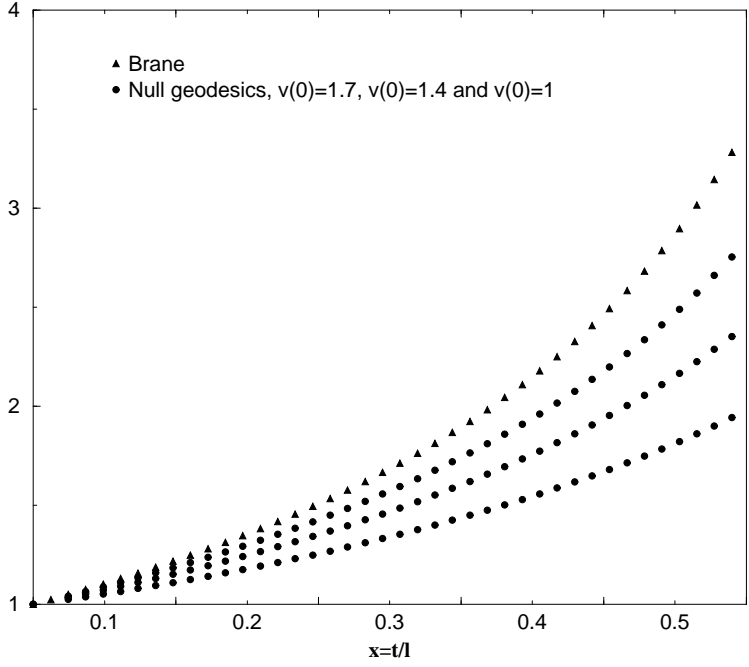}}\nonumber
\end{eqnarray}
\caption{The trajectory of the brane in the bulk for 
$\Omega_0=2$. Other curves are geodesics that start in the brane with
different initial velocities.}
\label{graphprimitivo}
\end{center}
\end{figure*}

Before the critical time, from the point of view of the bulk, there
is no time left for the remaining graviton geodesics to reach the
brane. This is in fact the kind of behaviour that we find in numerical
studies of the complete equation for the evolution of the brane and null geodesics as in the example of figure
\ref{graphprimitivo}.

\begin{figure*}[htb!]
\vspace{-2cm}
\begin{center}
\leavevmode
\begin{eqnarray}
\epsfxsize= 9.0truecm \rotatebox{-90}
{\epsfbox{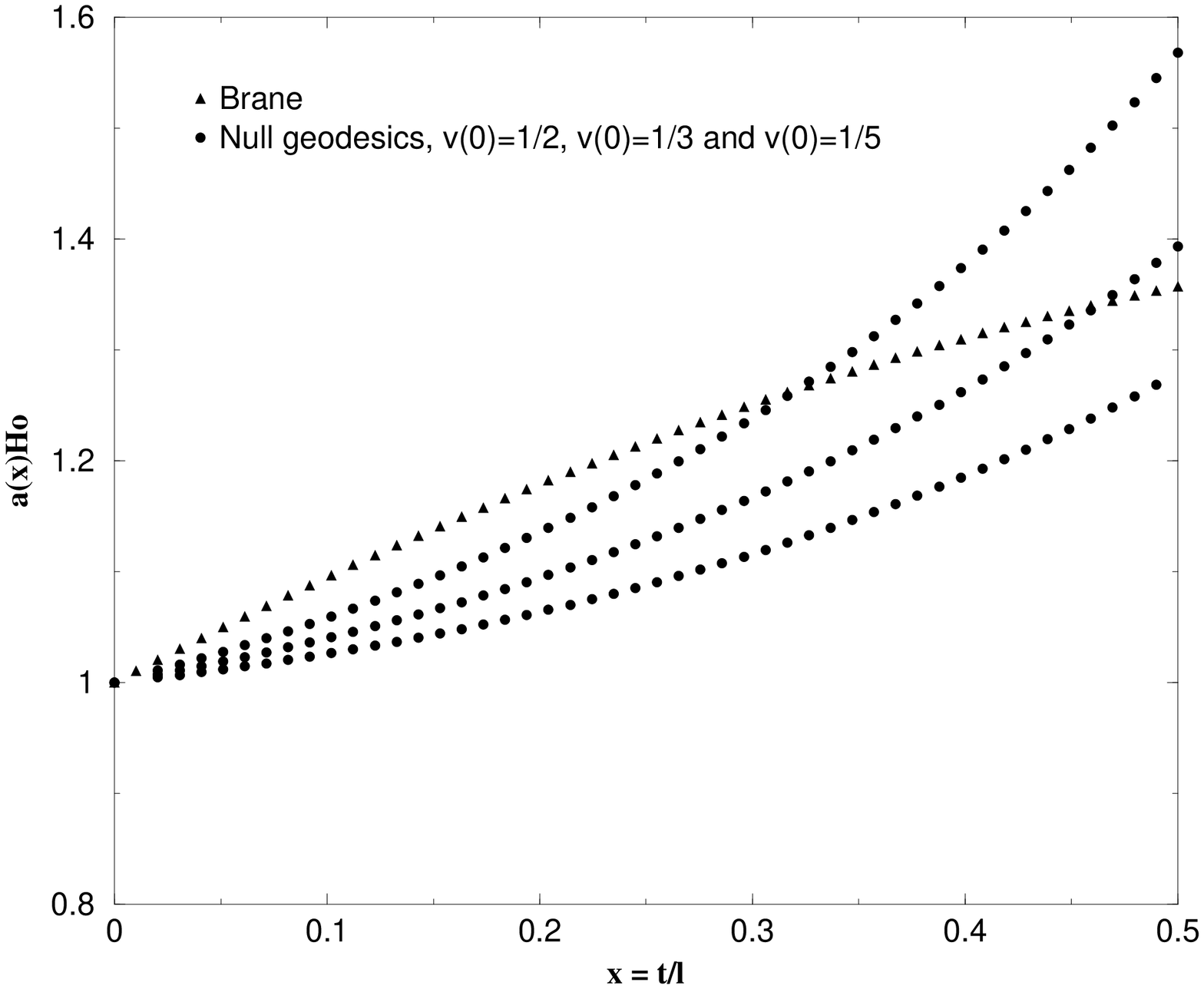}}\nonumber
\end{eqnarray}
\caption{The trajectory of the brane in the bulk for $\Omega_0=2$, matter 
dominated era (MDE) with $k=1$. Also null geodesics starting in
the brane with various initial velocities.}
\label{atual}
\end{center}
\end{figure*}

On the other hand, for later times the evolution of the brane is softer,
and shortcuts appear, as exhibited in fig. \ref{atual}.

\newpage

\subsection{The Effect of Shortcuts in Late Times Universes}

The shortcuts just found for late times Universes could be used to probe
the extra-dimensionality by the aparent violation of causality in the
brane. Thus, suppose that in $\tau=t=0$ an object could emit eletromagnetic and
gravitational waves and that we could be able to detect both signs in
times $\tau'_{\gamma}$ and $\tau'_{g}$. 
We compute now the order of magnitude of the advance in time of the
graviton. Since the signals cover the same distance in the brane,
\be
\int_{0}^{\tau'_{\gamma}} \f{d\tau_{\gamma}}{a_{b}(\tau_{\gamma})} = \int_{0}^{t'_{g}}\f{dt_{g}}{a(t_{g})}
\sqrt{h(a) - \f{\dot{a}(t_g)^{2}}{h(a)}}\quad . \label{intgeo}
\ee
Here,  $a$ denotes the coordinate defining the geodesic in the bulk, and
differs from the coordinate of the brane $a_b$. In terms of the
dimensionless parameters $y$ and $x$, $a=Ly$ and $t=Tx$ the last integral
reads 
\bear
\int_{0}^{t'_{g}} \f{dt_{g}}{a(t_{g})}\sqrt{h(a) -
\f{\dot{a}(t)^{2}}{h(a)}} &=& \int_{0}^{t'_{g}} \f{dt_{g}}{l}\sqrt{1 + \f{l^{2}}{L^{2}y^{2}}  -
\f{l^{2}}{T^2}\f{\dot{y}(x)^{2}}{y^{2} + y^{4}\f{L^{2}}{l^{2}}}}\quad .
\eear
Using the relation between the time in the brane and that of the bulk we
convert the above expressions in terms of the interval of time of the
observer in the brane, 
\bear
\int_{0}^{\tau'_{\gamma}}\!\!  \f{d\tau_{\gamma}}{a_{b}(\tau_{\gamma})}\!\! 
&=&\!\! \int_{0}^{\tau'_{g}} \f{d\tau_{g}}{l}\f{d
t_{g}}{d\tau_{g}}\sqrt{1 + \f{l^{2}}{L^{2}y^{2}}  -
\f{l^{2}}{T^2}\f{\dot{y}(x)^{2}}{y^{2} + y^{4}\f{L^{2}}{l^{2}}}}
\nonumber\\ 
 &=& \int_{0}^{\tau'_{g}} \f{d\tau_{g}}{l}\f{1}{h(a_{b})}\sqrt{h(a_{b}) + \dot{a}_{b}
(\tau_g)^2}\sqrt{1 + \f{l^{2}}{L^{2}y^{2}}  - \f{l^{2}}{T^2}\f{\dot{y}
(x)^{2}}{y^{2} + y^{4}\f{L^{2}}{l^{2}}}}\, .\nonumber \label{aux11}
\eear

The Friedmann equation implies
\bearst
&&\int_{0}^{\tau'_{g}} \f{d\tau_{g}}{l}\f{1}{h(a_{b})}\sqrt{\f{y_{b}^{2}L^{2}}{l^{2}}+
\Lambda_4L^2y_b^2 +
\f{\Omega_{0}}{y_{b}^{q-2}}\f{L_{c}^{q}}{l^2L^{q-2}}}\sqrt{1 +
\f{l^{2}}{L^{2}y^{2}}  - \f{l^{2}}{T^2}\f{\dot{y}(x)^{2}}{y^{2} +
y^{4}\f{L^{2}}{l^{2}}}}  \\ 
&&\quad = \int_{0}^{\tau'_{g}} \f{d\tau_{g}}{l}\f{1}{h(a_{b})}\f{y_{b}L}{l}\sqrt{1 +
l^2\Lambda_4+ \f{\Omega_{0}}{y_{b}^{q}}\f{L_{c}^{q}}{L^{q}}}\sqrt{1 +
\f{l^{2}}{L^{2}y^{2}}  - \f{l^{2}}{T^2}\f{\dot{y}(x)^{2}}{y^{2} +
y^{4}\f{L^{2}}{l^{2}}}} \\ 
&&\quad =\int_{0}^{\tau'_{g}} \f{d\tau_{g}}{a_{b}(\tau_{g})}\f{1}{1  +
\f{l^{2}}{L^{2}y_{b}^{2}}}\sqrt{1 +l^2\Lambda_4 +
\f{\Omega_{0}}{y_{b}^{q}}\f{L_{c}^{q}}{L^{q}}}\sqrt{1 +
\f{l^{2}}{L^{2}y^{2}}  - \f{l^{2}}{T^2}\f{\dot{y}(x)^{2}}{y^{2} +
y^{4}\f{L^{2}}{l^{2}}}}\quad .
\eearst

Therefore, if $L>>L_{c}>>l$ we obtain, at second order in  $L_{c}/L$ and $l/L$,
\[
\int_{0}^{\tau'_{g}} \f{d\tau_{g}}{a_{b}(\tau_{g})}\Bigl(1 -
\f{l^{2}}{L^{2}}\f{1}{y_{b}^{2}}\Bigr)\Bigl(1 +\f{1}{2}l^2\Lambda_4 +
\f{1}{2}\f{\Omega_{0}}{y_{b}^{q}}\f{L_{c}^{q}}{L^{q}}\Bigr)\Bigl(1 +
\f{1}{2}\f{l^{2}}{L^{2}y^{2}}  -
\f{1}{2}\f{l^{2}}{T^2}\f{\dot{y}(x)^{2}}{y^{2} +
y^{4}\f{L^{2}}{l^{2}}}\Bigr)\quad .
\]

Finally,
\bearst
\int_{0}^{\tau'_{\gamma}} \f{d\tau_{\gamma}}{a_{b}(\tau_{\gamma})} &=& 
\int_{0}^{\tau'_{g}} \f{d\tau_{g}}{a_{b}(\tau_{g})}\Bigr[1  + \f{1}{2}\f{\Omega_{0}}
{y_{b}^{q}}\f{L_{c}^{q}}{L^{q}}  + \f{1}{2}l^2\Lambda_4- \f{l^{2}}{L^{2}}
\f{1}{y_{b}^{2}} \nonumber \\ &+& \f{1}{2}\f{l^{2}}{L^{2}y^{2}}  -
\f{1}{2}\f{l^{4}}{T^2L^2}\f{\dot{y}(x)^{2}}{y^{4}}\Bigl]\quad .
\eearst

Thus, at first order, the time difference between the photon and the
graviton is corrected in the integrand by terms of the order $ \f{L_{c}^{q}}{a_{b}^{q}}$.
Today this factor is at most $10^{-58}$ and in the time of
decoupling  $10^{-46}$, showing that
the time advance of the graviton can
be safely neglected and is of no physical significance, in spite of the 
fact that the trajectory of the brane is distinctively different from the 
null geodesic.

\subsection{The Effect of Shortcuts in Early Time Universe}

From the analysis developed so far we learned that the periods of
evolution of the Universe differ by the scale that defines
the physical significance of the shortcuts. In an era where $a_{b}<<L_{c}$, the
trajectory of the brane in the bulk differs from the extreme geodesics by
$(a_{b}/L_{c})^{2q}$ and the shortcuts do not appear, since the brane itself
provides the graviton geodesic.

In the period when $a_{b}>>L_{c}$, the trajectory of the brane is far from a
null geodesic and shortcuts appear, but they are not significant, since the
{\it skin depth} of the graviton in the bulk is defined by the parameter
$l<<L_{c}<<a_{b}$. The difference between the time intervals of the photon and
the graviton is of the order $(L_{c}/a_{b})^{q}$.

\begin{figure*}[htb!]
\vspace{-1cm}
\begin{center}
\leavevmode
\begin{eqnarray}
\epsfxsize= 10.0truecm \rotatebox{-90}
{\epsfbox{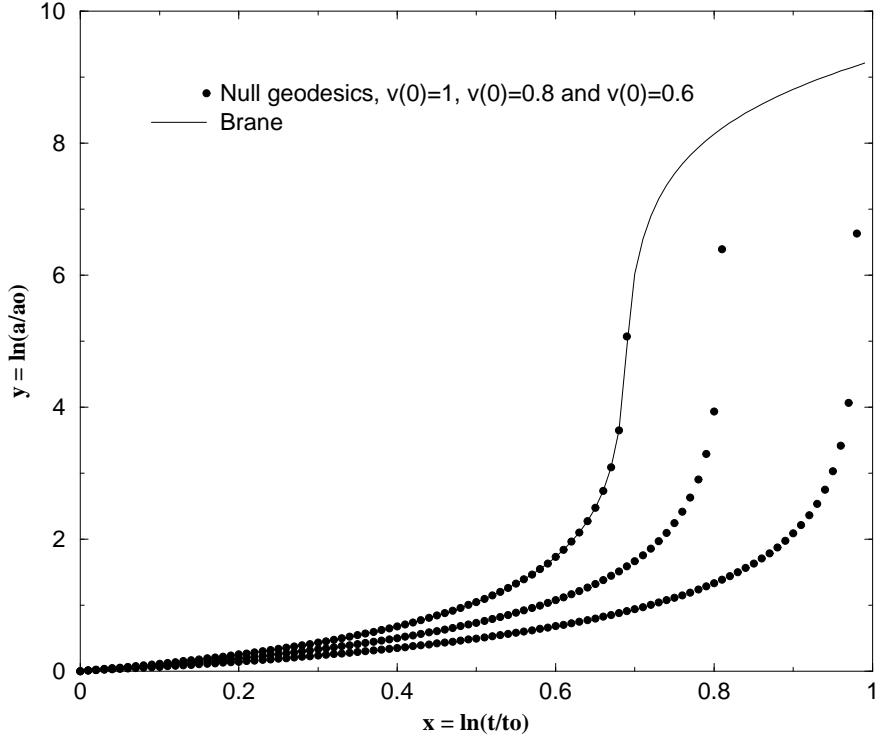}}\nonumber
\end{eqnarray}
\caption{The trajectory of the brane in the bulk leaving
$a=0.01 L_c$ in the   RDE. Geodesics in the bulk with different initial
velocities are exhibited intercepting the brane after the critical time
$t_c$.} 
\label{horizonte}
\end{center}
\end{figure*}

However, from the continuity of the evolution of the brane in the
bulk, we expect that there is also an intermediate situation
$a_{b}\sim L_{c}$
 when physically important shortcuts could appear, since the evolution of
the brane is far enough from a geodesic. Indeed, in figure  \ref{horizonte},  rescaling the geodesics in the bulk we can observe the
behaviour of the brane as compared to the geodesics that start in it at a
later time. It is clear that these shortcuts are
serious mediators of homogenization of the matter in the brane in the era
before nucleosynthesis \cite{Ishihara}, \cite{Caldwell}.

From the evolution of the brane in the bulk in the intermediate epoch we thus conclude that there is a critical age  $t_{c}$, after which the
gravitational waves leaving the brane return before the arrival of the
photons released at the same time as the gravitons. The behaviour of
the geodesic in the bulk
shows that any geodesic starting in the brane at a certain instant will be singular at a time
later than the critical time, indicating that it will return to the
brane. Thus, information leaks between regions which apparently are causally
disconnected at times $t>t_{c}$.

In order to study the horizon problem, we now compare, at a certain time $\tau_{n}$ previous to
nucleosynthesis, the graviton horizon $R_{g}$ with the
observable proper distance of the universe (from the radiation
decoupling
until today) \cite{Chung}. Since the graviton evolves in a bulk geodesic,
\bear
R_{g} &\equiv& \int \f{dr}{\sqrt{1-kr^{2}}}  = \int \f{dt}{a} \sqrt{h(a) -
\f{\dot{a}^{2}}{h(a)}} = \int \f{d\tau}{a}\f{\sqrt{h(a_{b}) + \dot{a}_{b}^{2}}}{h(a_{b})} \sqrt{h(a) -
\f{\dot{a}^{2}}{h(a)}}\quad \nonumber \\  &=&  
\int_0^{\tau_{n}} \f{d\tau}{a_b}\f{1}{1 + l^{2}/a_b^2}\sqrt{1 +
\Lambda_4 l^2 + \f{\Omega_0L_c^q}{a_b^q}\Bigl(1 +
\f{\Omega_0L_c^q}{4a_b^q(1+\Lambda_4l^2)}\Bigr)}\times\nonumber\\
&&\times\sqrt{1 + \f{l^{2}}{L^{2}y^{2}} - \f{l^{2}}{T^2}\f{\dot{y}(x)^{2}}
{y^{2} + y^{4}\f{L^{2}}{l^{2}}}} \quad ,\label{graviton1}
\eear
while in the brane, the size of the observable Universe is
$R =  \int \f{d\tau}{a_b(\tau)}$.

We use the known results for the usual particle horizon
\[
R  = \f{1}{\sqrt{\Omega_0}H_0a_{b0}}\int_{z(\tau)}^{z(0)}z^{-q/2}dz
=\f{2}{(q-2)\sqrt{\Omega_0}H_0a_{b0}} (z(\tau)^{1-q/2} -  z(0)^{1-q/2})\quad.
\]
where $z(\tau) = a_{b0}/a_{b}(\tau)$. Today, 
\[
R \sim \f{2}{\sqrt{\Omega_0}H_0a_{b0}} \equiv R_0\quad.
\]

For the computation of the graviton horizon, we work in a primordial era
before nucleosynthesis. Thus, in the Friedmann equation we can neglect the
usual cosmological term as well as the curvature term. At an epoch between Planck era and nucleosynthesis,
$l<< a_b << H_0$, it is  safe to neglect terms involving  $l^2/L^2$ in
(\ref{graviton1}), and we find
\[
R_g \approx 
\int_0^{\tau_{n}} \f{d\tau}{a_b}\sqrt{1 + \f{\Omega_0L_c^q}{a_b^q} +
\f{\Omega_0^2L_c^{2q}}{4a_b^{2q}}}\quad .
\]

Using the Friedmann equation we get
\be
R_g = 
\f{1}{\Omega_{0}^{1/2}H_{0}a_{b0}}\int_{z(\tau_n)}^{z(0)} \f{dz}{z^2} \f{\Bigr(1 + \f{\Omega_0H_0^2l^2}{2}z^4\Bigl)}{\sqrt{ 1 + 
\f{\Omega_0 H_0^2l^2}{4}z^4}}\quad.
\ee

The integral diverges for arbitrarly high redshifts, proving that the
horizon problem is potentially solvable. The behaviour of this integral
can be determined. We have
\[
R_g \approx 
\f{l}{a_{b0}}z(0)\quad.
\]

Comparing with the size of the Universe today,
\[
\f{R_g}{R_0}  \sim \f{\sqrt{\Omega_0}}{2}H_0l z(0)\quad.
\]

It looks like shortcuts are not enough to solve the horizon problem, since
we would need to go back in time to $z(0) \sim (H_0l)^{-1} \sim
10^{29}$, $10^{11}$ times higher than the redshift at the Planck time
in the brane associated with the fundamental scale of gravity,
$\kappa_{5}$. 

For the time being, lets assume the validity of the theory in such
high energies. In the next section we will argue that, in order to
provide a solution to the usual cosmological problems, inflationary
models must use of those high redshifts. We may note, however, that there are actually two
related time scales. 
In the primordial  Universe the brane is evolving as a part of the bulk, with velocities 
close to that of light, and time intervals  in the brane correspond to 
much longer intervals in the bulk. In fact, the relation between these 
scales is obtained from $t_{0}=\tau_{0}=a_{b}(\tau_0)=0$ and from
\[
t  = \int_{0}^{\tau}d\tau
\f{\sqrt{h(a_{b})+\dot{a}_{b}^{2}(\tau)}}{h(a_{b})}\quad . 
\]

For $a_{b}<<L_{c}$ we have $\tau<<l$, and one finds, as a consequence
\[
t  \approx \int_{0}^{\tau}d\tau \f{\dot{a}_{b}(\tau)}{h(a_{b})} =
\int_{0}^{a_{b}(\tau)}\f{da_{b}}{k+a_{b}^{2}/l^{2}}\quad .
\]
This implies, for the example of a closed universe with $\Omega_0=2$,
\[
t = l\arctan\Bigl(\f{a_{b}(\tau)}{l}\Bigr)\quad .
\]

Therefore, when $a_{b}(\tau)\sim l$, that is, when the brane time is 
$\tau \sim 10^{-67}s$, the corresponding bulk time $t\sim \f{\pi}{4}l \sim
10^{-11}$s is much larger than the Planck scale, $t >> M_{(5)}^{-1}$.

If we assume that quantization is mandatory according to the bulk Planck
energy scale, geodesics that start in the bulk with $z(0)\sim (H_0l)^{-1}$
should be sufficient to homogenize the Universe before nucleosynthesis,
reaching  $R_{g}\sim 1$. In order to verify this, we have
numerically integrated the expression for the distance reached by
geodesics 
$a(t)$ starting in the brane at $a_{b}=l$, corresponding to a time
$t=\pi/4 l$, for $k=1$ with
$\Omega_0\sim 2$ ($a_{b0}\sim H_0^{-1}$):
\be
R_{g} = \int_{l\pi/4 }^{t_f} \f{dt}{a} \sqrt{h(a) -
\f{\dot{a}^{2}}{h(a)}}\quad .
\label{acima} 
\ee

\begin{figure*}[htb!]
\vspace{-1cm}
\begin{center}
\leavevmode
\begin{eqnarray}
\epsfxsize= 8.0truecm \rotatebox{-90}
{\epsfbox{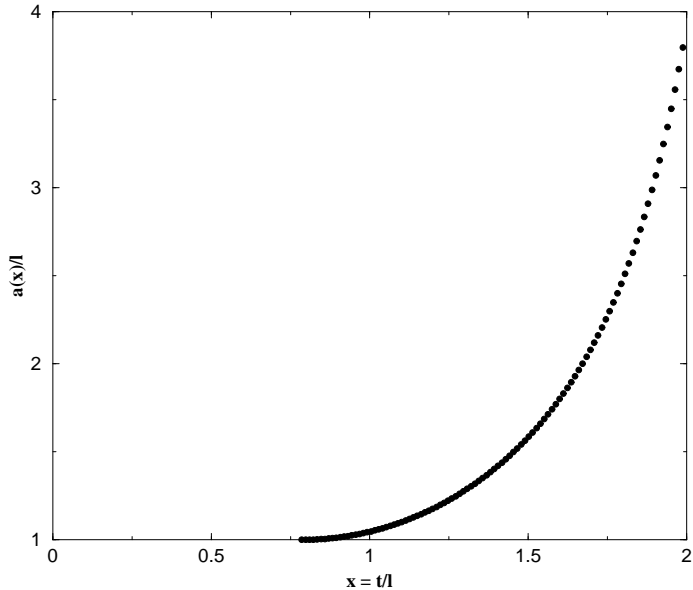}}\nonumber
\end{eqnarray}
\caption{Null geodesic starting at $a=l$, $t=\pi/4 l$ with initial velocity
$v(\pi/4)=0$.}
\label{horizonte3}
\end{center}
\end{figure*}

The solution of the geodesic equation is given in figure
\ref{horizonte3}. The brane evolves initially as a geodesic with speed
$v_b\sim 2$ ($a_{b}\sim l$) and we exhibit a geodesic with zero initial
speed in the bulk. The point of inflexion in its trajectory is  $t_{f} =
2.356 l > t_{c} = \f{\pi}{2}l$. As we saw before, the brane evolves
smoothly after the transition and the geodesics return to the brane if 
we wait  until $t_{f}$, see (\ref{acima}). Although the trajectory 
diverges near this point, the integrand in  (\ref{acima}) is finite and 
we can  perform the integration, finding
\[ 
R_{g} = 2.526 \quad .
\]

Thus, the observable Universe is
smaller than the reach of the relic gravitons that start in a previous
era.

\section{Causal Gravitational Structure}

In the last  section we learned about the  behaviour
of null-geodesics for closed Universes. We can, however,  study analytically 
the whole causal structure of the gravitational signs for a $k=0$ Universe.

The geodesic equation, (\ref{nullgeodesic}), is quite simple for
purely AdS $k=0$ space-times
\be
\f{1}{a^{2}}\f{d a}{dt}  = \f{v(0)}{a(0)^{2}} \quad.
\ee

Thus, a geodesic that starts in the brane at $a=a(0)$ ant $t=0$, with
initial veolcity $v(0)$, returns to it when
\[
t_r \approx \f{a(0)}{v(0)} \quad.
\]

The expression for the gravitational horizon can also be integrated
\bear
R_{g} &=& \int_{0}^{t_{r}}\f{dt}{a}\sqrt{h(a) - \f{\dot{a}^{2}}{h(a)}}
= \int_{0}^{t_{r}}\f{dt}{l}\sqrt{1 - \f{l^{4}\dot{a}^{2}}{a^{4}}}
\nonumber \\ &=& \f{t_{r}}{l}\sqrt{1 - \f{v(0)^{2}l^4}{a(0)^{4}}}\quad.
\eear
Using the relation between the returning time and the initial velocity,
\bear
R_{g} = \f{t_{r}}{l}\sqrt{1 - \f{l^{4}}{a(0)^{2}t_r^{2}}}\quad.
\eear

In order to relate the returning time  to  the
redshift,  we must integrate the relation between time of the
bulk and time of the brane (\ref{timetime}). We already know that
from $a(0)$ to the critical period of transition a time of $t_c\sim
\f{l^{2}}{a(0)}$ has passed. After that, the evolution is dominated by the usual
term in Friedmann equation and we can approximate
\bear
t_{r} &\approx& \f{l^2}{a(0)} + \int_{L_{c}}^{a_r}l\f{da_b}{a_b^{2}H}
\approx \f{l^2}{a(0)} +
\int_{L_{c}}^{a_r}da_b\f{l}{\sqrt{\Omega_0}H_{0}}a_{b}^{q/2-2}
\nonumber \\ &\approx& \f{l^2}{a(0)} +
\f{2l}{(q-2)\sqrt{\Omega_0}H_{0}a_{b0}}\Bigl(z_{r}^{-q/2+1} - 10^{-15}\Bigr)\quad,
\eear
where $z_r$ must, of course, be greater than the redshift in the
transition, $z_{Lc}\sim 10^{15}$. 

Substituting back in the expression of the gravitational horizon,
\bear
R_{g} &=& \Bigl[\f{l}{a(0)} +
\f{2}{(q-2)\sqrt{\Omega_0}H_{0}a_{b0}}\Bigl(z_{r}^{-q/2+1} -
10^{-15}\Bigr)\Bigr]\nonumber\\ &\times& \sqrt{1 -  \Bigl[1 + \f{2a(0)}{(q-2)\sqrt{\Omega_0}lH_{0}a_{b0}}\Bigl(z_{r}^{-q/2+1} - 10^{-15}\Bigr)\Bigr]^{-2} }\quad.
\eear

When considering the interesting situation of a high initial redshift $z(0) =
a_{b0}/a(0) > (H_0l)^{-1}$, this expression can be approximated by
\bear
R_{g} \approx \f{l}{a(0)}\sqrt{\f{4a(0)}{(q-2)\sqrt{\Omega_0}lH_{0}a_{b0}}\Bigl(z_{r}^{-q/2+1} - 10^{-15}\Bigr) }\quad.
\eear

Comparing with the size of the horizon today, we find
\bear
\Bigr(\f{R_{g}}{R_{0}}\Bigl)_{z_r} \approx
\sqrt{\sqrt{\Omega_0}\f{lH_0}{(q-2)}}\sqrt{z(0)\Bigl(z_{r}^{-q/2+1} - 10^{-15}\Bigr)} \quad.
\label{completehorizon}
\eear

We note, as we did in the last section, that if sufficiently large
redshifts were available, the graviton horizon in a past epoch could be
larger than the present size of the observable Universe.

We argue, however, that those high redshifts could be available
even when the energy density in the brane is not quantized
yet. Indeed, if inflation takes place in the brane high redshifts
could be present in the beginning of the
inflationary epoch.

 Denoting the redshift when
inflation  ends by $z_e$, if the size of the present Universe, $R_0$, is
expected to be in causal contact during inflation, we must reach at
least a redshift in the beginning of inflation, $z(0)$, that solves
\[
\f{a_{b0}R_{0}}{z(0)} = H^{-1}(z_{e}) \quad.
\]

The unusual results in brane-world cosmology are expected if inflation
ends before the transition time, when the quadratic term in Friedmann
equation dominates. In this case,  we get
\[
\f{a_{b0}R_{0}}{z(0)} = \f{1}{\Omega_{0}H_{0}^{2}l z_e^{4}}
\]
and 
\be
z(0)  =  2\sqrt{\Omega_{0}}H_{0}l z_e^{4}\quad. 
\label{result}
\ee

If $H(z_e)l>>1$, it is simple to
note, from equation (\ref{timetime}), that the evolution of the brane
in the bulk is not altered during inflation. Thus, we can substitute the result(\ref{result}) in the complete expression for the causal
gravitational horizon (\ref{completehorizon}),
\bear
\Bigr(\f{R_{g}}{R_{0}}\Bigl)_{z_r} \approx
\sqrt{\f{2\Omega_0}{(q-2)}} H_{0}l z_e^{2}\sqrt{\Bigl(z_{r}^{-q/2 + 1} - 10^{-15}\Bigr)} \quad.
\eear

This equation tells us that,  in the time of
nucleosyntesis, when $z_r\sim 10^{10}$,  $R_{g}/R_{0}\sim 1$ for a model with inflation ending just
before what would be the Planck era, with $z_e \sim 10^{17}$ (where
$z_{Pl} \sim 10^{18}$). This proves that successfully inflationary
models ending before the transition time necessarily  make relevant changes in the causal structure of
the universe. 

\begin{figure*}[htb!]
\vspace{-1cm}
\begin{center}
\leavevmode
\begin{eqnarray}
\epsfxsize= 10.0truecm \rotatebox{-90}
{\epsfbox{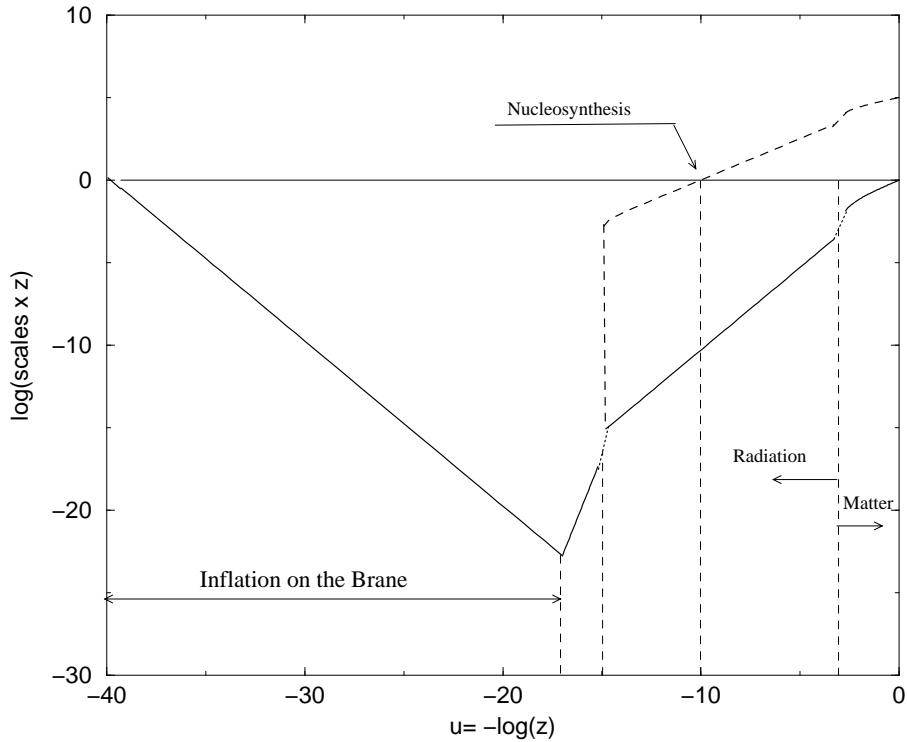}}\nonumber
\end{eqnarray}
\caption{We plot in log-log scale the evolution of the fraction of
graviton horizon and the present observable size,
$R_{g}/R_0$ (dashed line) and the same for the Hubble horizon
$a_{b}^{-1}H^{-1}/R_{0}$ (solid line), for an inflationary model in the brane that ends after Planck
time ($u_{PL}=-\log(z_{PL}) = -18$),  and before transition time
($u_{TR} = -15$),  in $u_{end} = -\log(z_{end}) = -17$. The present
scale would be under the de-Sitter horizon if redshifts like
$u(0)=-39$ were available. This imply in strong modifications of the
causal structure for gravitational signs after the transition time.}
\label{inflation}
\end{center}
\end{figure*}

We are able to sketch the gravitational horizon for this kind of
configuration. In figure \ref{inflation} we show the behaviour of the
Hubble horizon in comoving coordinates $H^{-1}a_{b}^{-1}/R_{0}$ and the graviton horizon $R_g/R_0$ for an
inflationary model ending just before the Planck era, $z_e=10^{17}$ and
producing the necessary number of e-folds to solve the horizon
problem. The fraction of the Universe in causal
contact by gravitational signs in the nucleosynthesis epoch is just
the present horizon $R_{0}$. Today, the gravitational horizon would be
$10^{5}R_0$.

\section{Conclusions}

Studying the shortcut problem in braneworld cosmology from the point of
view of the bulk we have explicitly shown that shortcuts are indeed
common in late time Universes, though they are extremelly small and the time advance of the graviton can
be safely neglected. However, we have also learned that gravitational signs may  leave and
subsequently return to the brane even in early times Universes. We
have showed that those shortcuts
exist  and that the new  scale of the model, $l$, implies in a
minimum time scale for the reception of those signs by an observer in the brane. Before
that critical time, the brane itself evolves like a null-geodesic in the bulk.

If high initial redshits were available the shortcuts just found could
solve the horizon problem without inflation. More important
however,  may be the
effect of those shortcuts with an inflationary epoch in the brane.
   
 Brane-world  models incorporate two
changes in the cosmology, namely, the modified Friedmann equation and
the possibility of leaking of gravity in the extra dimension. Using
the first of those modifications, it was shown that, remarkably, the
consistency equation is mantained in the brane-world formalism when
the inflation is guided by a scalar field minimally coupled in the
brane \cite{inflation}. This consists in bad news for those who expect that brane
cosmological configurations could probe the extra dimensionality of
our Universe.

However, we have showed that if there was an
inflationary epoch in the brane evolution, the causal
structure of the universe could be strongly modified. This could be a
sign of an unusual evolution of the perturbations  from the
time they cross the de Sitter horizon, $H^{-1}$, during inflation,
through the time they became causally connected again. In this case,
there could be  distinct
predictions for the  microwave background radiation structure even with the same
consistency equation  during inflation. 
Thus, further investigation on the dynamics of perturbations in
inflationary brane-world models may prove
useful to probe the dimensionality of space-time.

\bigskip
{\bf Acknowledgements:} This work has been supported by Funda\c c\~ao
de Amparo \`a Pesquisa do Estado de S\~ao Paulo {\bf (FAPESP)} and Conselho
Nacional de Desenvolvimento Cient\'\i fico e Tecnol\'ogico {\bf (CNPq)}, 
Brazil. We would like to thank R. Abramo for reading the manuscript.


\end{document}